# TER: a robot for remote ultrasonic examination. Experimental evaluations

J.J. Banihachemi[1,2] (MD, MSc), E. Boidard[2], J. L. Bosson[2,3] (MD, PhD), L. Bressollette[4,8] (MD), I. Bricault[2,5] (MD, PhD), P. Cinquin[2,6] (MD, PhD), G. Ferretti[5] (MD, PhD), M. Marchal[2] (MSc), T. Martinelli[5] (MD), A. Moreau-Gaudry[2,3,6] (MD, PhD), F. Pellissier[7], C. Roux[8] (PhD), D. Saragaglia[1] (MD), P. Thorel[7] (PhD), J. Troccaz[2] (PhD), A. Vilchis[2,*] (PhD)

[1]Emergency Department, Grenoble University Hospital (South), Grenoble, France
[2]TIMC Laboratory, Grenoble, France
[3]Clinical Investigation Centre, Grenoble University Hospital, France
[4]Vascular Medicine Department, Brest Hospital, France
[5]Radiology Department, Grenoble University Hospital (North)
[6]Technological Innovation Centre (CIT), Grenoble University Hospital
[7]France Telecom R&D
[8]LATIM Laboratory, Brest, France
[*]A. Vilchis is now with the UAEM University, Mexico

Authors for correspondence: Jocelyne.Troccaz@imag.fr or Philippe.Cinquin@imag.fr
IN3S – School of Medicine – Domaine de la Merci – 38706 La Tronche cedex – France

This chapter:
- Motivates the clinical use of robotic tele-echography
- Introduces the TER system
- Describes technical and clinical evaluations performed with TER.

## 1. Introduction

Ultrasonic (US) imaging is very frequently used for diagnostic or interventional procedures. This widespread modality requires a very good expertise of the operator, depending on the anatomical region to be explored and/or on the pathology type. Such an expert is not always available close to the patient. Because patient transportation to a suitable site may not be feasible or desirable, telemedicine may be very useful. However, contrarily to other imaging modalities, the only transmission of series of ultrasonic images to remote experts for later interpretation is not satisfactory. US image acquisition itself is really part of the expertise and conditions the clinician ability to interpret the images. Indeed, the way the clinician moves the US probe on the patient body directly contributes to his/her understanding of series of images. This is why several groups, worldwide, explored over the last decade robotic tele-ultrasound: the objective is to develop robotic systems that move the US probe on the patient's body and to give the control of those motions to the expert operator located in another site. In that way, the remote expert both acquires images and interprets them. Several prototype systems have been developed [1,2,3,4,5] but as far as we know very few clinical evaluations were done. References [6,7] respectively report a first global feasibility analysis of cardiac and abdominal US exploration for 20 patients and a clinical evaluation of remote examination of 29 pregnant women located in an isolated hospital of North Morocco. Our objective is to contribute to the clinical demonstration of robotic tele-echography. In this paper we introduce the TER system and describe technical and clinical experiments.

## 2. The TER system

### 2.1 System architecture

TER is composed of two subsystems placed on two distant sites: the master site where the expert is located and the slave site where the patient is. The master subsystem is composed of a haptic device connected to a computer workstation. This device, a Phantom® from Sensable Devices Inc., is manipulated by the expert as a virtual ultrasound probe. In this way he/she remotely controls the real probe based on the US images he/she receives from the slave site and with the help of a force feedback enabling a fine control of the pressure exerted by the real probe onto the patient's body. The slave subsystem is composed of the ultrasound system and of a robot both connected to a second computer workstation. The robot carries the real US probe and enables moving it on the patient's body. Both subsystems include audio-video teleconferencing facilities. In this way, the expert can see the patient and can interact with him/her and with the staff person located at the slave site. This staff person (typically a nurse) helps for the patient installation and system set-up. The two subsystems are connected by a telecommunication link enabling transmission of the probe motion orders from the master site to the slave site, transmission of the US images, force information and robot parameters from the slave site to the master site and bi-directional transmission of audio and video data. This is summarized in figure 1.

*2.2 TER releases*
Two versions of the TER system were developed based on two slightly different robot architectures. TER-V1 [8] included a robot actuated by pneumatic muscles. Because control of the muscles turned out to be quite difficult and inaccurate, a second version of the robot was designed. TER-V2 [9,10] is electrically actuated. A similar philosophy underlies both designs: having a light and safe robot that can be positioned directly on the patient's body (typically on the abdomen) and is able to automatically adapt to different patient sizes and to motions of the body surface due to breathing for instance. For this reason, both robots are composed of a first substructure with four flexible straps that allow translation movements of the second substructure on the abdominal surface; this corresponds to two degrees of freedom (dof). The second substructure (four dof) is dedicated to rotational movements of the US probe and to fine translational motions ensuring a correct contact of the probe on the body. TER-V1 (respectively V2) implements this substructure by a parallel (respectively serial) architecture. TER-V2 robot can be seen on figures 1 and 2.

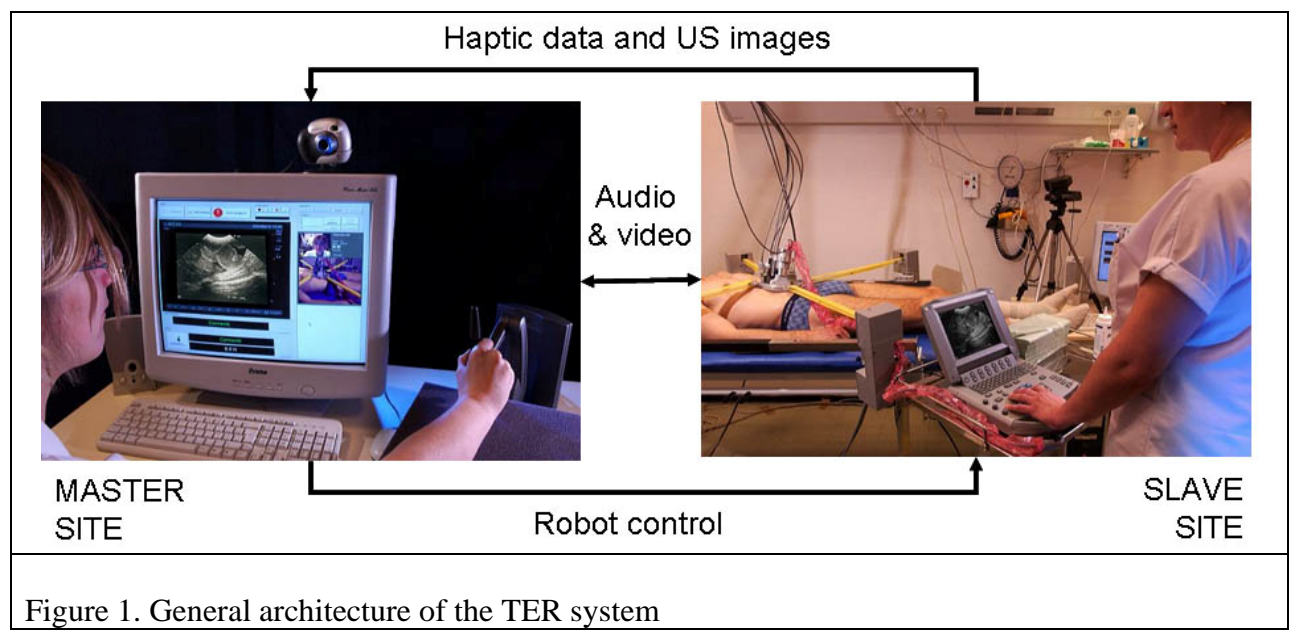

Figure 1. General architecture of the TER system

The Phantom® Desktop haptic device renders 3D force information in a 160x120x120mm workspace. Because the most important force component for probe control is the reaction force normal to the body surface, we alternatively developed a 1D force rendering system [11]. This haptic device has also the advantage of being a freehand system that gives the operator a much larger workspace than the Phantom®. This original haptic device has not yet been integrated to the TER prototypes.

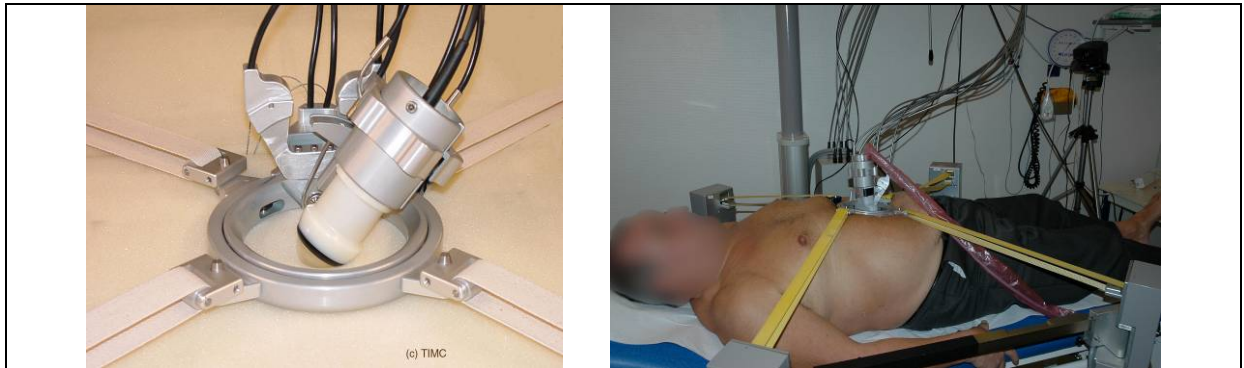

Figure 2. TER-V2 robot. (Right) The end-effector is translated by four straps (2dof); it orients and applies fine translations (4dof) to the US probe. (Left) In a real exam situation.

*2.3 Experimental evaluations*
Experimental work began by using TER-V1 for remote examinations of anatomical phantoms and then of a healthy volunteer. One radiologist and two gynaecologists participated to this first stage. For the first in-vivo experiment, the volunteer was located about 10km far from the expert operator. Two ISDN 128kb/s connections were used.

As regards TER-V2, first testing was performed using a LAN inside the TIMC laboratory. Then ISDN connections were used firstly for quite local testing (10km distance), then for longer distances. In October 2002, in the context of an exhibit, examinations were performed from physicians located in Toulouse on an anatomic phantom placed in Grenoble (600km distance). An ISDN 256kb/s was dedicated to the experiment (128kb/s for the US images – 4 images per second, 64kb/s for the audio-video teleconferencing data and 64kb/s for the haptic and robot data). In December 2002, in the context of a French military telemedicine congress, a volunteer located in Grenoble was examined by physicians located in Toulon (300km distance). An ISDN 512kb/s enabled transmission of the US images (320kb/s), audio-video data (128kb/s) and haptic and robot data (64kb/s). All those experiments were successful and physicians were enthusiastic about TER potential capabilities for real clinical situations.

## 3. Clinical experiments
The clinical experiments concern TER-V2 that we will name TER, for short, in the following.

*3.1 Clinical feasibility for angiology application*
This first trial was intended to evaluate the ability to practice US examinations remotely in a real clinical context. Compared to previous experimental work, the telecommunication network was an experimental one: the 10Gb/s VTHD network connecting the Grenoble hospital to the Brest Hospital (1125km distance). The TER prototype was ready for evaluation in June 2003. Preliminary experiments showed that the architectural choices, controllability and compatibility with the VTHD network were very satisfactory. In May and June, 4 volunteers participated to preliminary testing. Based on the positive feedback from the radiologists and from the volunteers, the experimental work with patients started. It concerned patients suffering from Aortic Abdominal Aneurysms (AAA) and/or from atherosclerosis. From July 2004 to march 2005, 58 patients were included by Grenoble and Brest hospitals.

All patients signed the consent form after being informed about the aim of the research and content of the protocol, in conformity with the French bio-ethic law.

Two complete systems were installed both in Brest and Grenoble enabling each site to be the master or the slave site. Both sites used the same US device for examination of the patients. Each included patient was examined both in the traditional way at his recruitment site – this initial exam was considered as being the reference one – and remotely using TER from the other site as concerns US control. Apart from the two involved physicians, another person (nurse, resident or MD) was present during the TER exam, at the slave site, in order to help installing the patient, setting-up the robot and using the US system (freezing the image, taking measurements, tuning image acquisition parameters). This person had to know how to use both the US device and the TER system, but he/she was not authorized to give any medical advice or comment during the exam.

Several pieces of information were recorded for both US exams (traditional/tele-robotic) and compared:
- Data concerning the pathology: presence of AAA or stage of atheromatosis and related measurements (for instance maximum antero-posterior diameter of the aneurism).
- Data related to the exam: the two first scores were given by the operator and concerned the feasibility and the global quality of the exam. The third score was the evaluation of the acceptability by the patient. The duration of the exam was also recorded.

The study demonstrated that remote examination of patients was feasible. The four failures were due to technical problems (2 dysfunctions of the Phantom®, 1 telecommunication link disruption and 1 computer crash). Exploration of the anatomical structures was possible and diagnostic elements were in agreement with the reference examinations. Remote measurements were repeatable and intra-observer variability was good in comparison with literature data. Remote exams were a little longer in average. Satisfaction rate of the physician was lower in average when performing the exam remotely whilst in the same time examinations are comparable as testify the other variables. The presence close to the patient is probably an important element for the physician feeling efficient. Patients' satisfaction rate was very good except for 2 of the 4 patients for which a technical problem occurred. Globally, the abdominal examinations of angiology pathologies with the TER system appeared feasible, safe and reliable.

*3.2 FAST versus TER in emergency trauma diagnosis*
The second clinical study takes place in the context of emergency trauma diagnosis. When echographic examination is required to detect the presence of an abdominal internal lesion, two techniques are feasible: one consists in having a complete US exam from a radiologist; the other one, called FAST (Focused Assessment Sonography for Trauma) is a specific examination performed by the emergency clinician. FAST being necessarily less sensitive than full US examination from a radiologist, false negative may occur. Since a radiologist may not be available at any time in trauma centres, one alternative would be to allow a radiologist to perform the examination remotely using TER. Thus, the study aims at comparing examinations performed by a radiologist using TER to examinations done by an emergency clinician using FAST.

The experiment takes place in Grenoble North and South Hospitals. The slave site is located in the Emergency Department of the South Hospital and the master site is installed in the

Radiology Department of the North Hospital. The two sites, distant of about 12km, are connected by the private Ethernet 100Mb/s link of the Hospitals network. In case of suspicion of a visceral trauma the patient has to be transferred from South hospital to North one for complementary examinations and specific care.

Seventy patients recruited by the emergency unit will be included in this study which started on May 2006, after authorization of the ethical committee. Each patient is examined both using TER and FAST in a random order. As for the angiology evaluation, different pieces of information are collected:
- diagnostic elements: in order to determine the agreement of the two examinations.
- Other data such as duration of the exam, satisfaction rate of the clinicians, acceptation rate of the patient, etc.

At the time of this publication, 11 patients have already participated to the study. For 10 of the 11 patients, no visceral trauma was detected either with FAST or with TER and this was confirmed by the patient follow-up. For the $11^{th}$ one, the TER examination was not satisfactory and the patient had to be transferred to the North hospital for direct examination by the radiologist. TER exams were in average longer that FAST ones (26mn versus 9mn). Of the 11 patients, 10 transfers to the North hospital were avoided resulting in direct cost reductions for those patients. Potential pain for trauma patients, due to the straps pressure on the abdomen, was not observed. Of course all planned patients have to be included before making any conclusion about the benefits of using TER for trauma diagnosis.

## 4. Conclusion

In this paper, we have shortly described the TER system and experiments aimed at demonstrating the technical and clinical feasabilities of the approach. Obtained results are quite encouraging but on-going work has to be finished before making any definitive analysis. Two clinical applications have been or are evaluated but those are obviously not the only two which are concerned by such an approach; the spectrum is potentially very large (patients at risk that cannot be transported, patients in very isolated places (including space, sea), elderly persons, etc.).

Industrial dissemination of such systems raises issues related to cost evaluation, economical model (which hospital pays what to whom?) and legal issues (in case of problems who is responsible for what?), etc. Those are complex questions that have not yet been satisfactorily answered and the fact that two countries may be involved with different regulations and laws could make it even more difficult.

**Acknowledgements**
This work has been supported by France Telecom R&D, Praxim-Medivision and by the French ministry of research (ACI Tele-medicine program and RNRT program).